\documentclass[aps, prl, twocolumn,showkeys,
showpacs,amsmath,amssymb,superscriptaddress,groupaddress]{revtex4-1}
\usepackage{graphicx}
\usepackage{dcolumn}
\usepackage{bm}
\usepackage{epsfig}
\usepackage{subfigure}
\usepackage{color}
\usepackage{epstopdf}

\usepackage{graphicx}  
\usepackage{dcolumn}   
\usepackage{bm}        
\usepackage{amssymb}   

\begin{document}
\title{Schramm-Loewner Evolution and isoheight lines of correlated landscapes}

\author{N. Pos\'e} \email{posen@ifb.baug.ethz.ch} \affiliation{ETH
  Z\"urich, Computational Physics for Engineering Materials, Institute
  for Building Materials, Wolfgang-Pauli-Strasse 27, HIT, CH-8093 Z\"urich,
  Switzerland}

\author{K. J. Schrenk } \email{kjs73@cam.ac.uk} \affiliation{Department of Chemistry, University of Cambridge, Lensfield Road, Cambridge CB2 1EW, UK}

\author{N. A. M. Ara\'ujo} \email{nmaraujo@fc.ul.pt} \affiliation{Departamento de F\'isica, Faculdade de Ci\^encias, Universidade de Lisboa, 1749-016 Lisboa, Portugal, and Centro de F\'isica Te\'orica e Computacional, Universidade de Lisboa, 1749-016 Lisboa, Portugal}

\author{H. J. Herrmann}\email{hjherrmann@ethz.ch} \affiliation{ETH
  Z\"urich, Computational Physics for Engineering Materials, Institute
  for Building Materials, Wolfgang-Pauli-Strasse 27, HIT, CH-8093 Z\"urich,
  Switzerland} \affiliation{Departamento de F\'isica, Universidade
  Federal do Cear\'a, Campus do Pici, 60455-760 Fortaleza, Cear\'a,
  Brazil}

\begin{abstract}
Real landscapes are usually characterized by long-range height-height
correlations, which are quantified by the Hurst exponent $H$. We analyze
the statistical properties of the isoheight lines for correlated
landscapes of $H\in [-1,1]$. We show numerically that, for $H\leq 0$ the
statistics of these lines is compatible with $SLE$ and that established
analytic results are recovered for $H=-1$ and $H=0$. This result
suggests that for negative $H$, in spite of the long-range nature of
correlations, the statistics of isolines is fully encoded in a Brownian
motion with a single parameter in the continuum limit. By contrast, for
positive $H$ we find that the one-dimensional time series encoding the
isoheight lines is not Markovian and therefore not consistent with
$SLE$.
\end{abstract}


\maketitle

We study isoheight lines of long-range correlated landscapes. They are
the paths of constant height in topography~\cite{Boffeta08}, the
equipotential lines on energy landscapes~\cite{Ziman68,Weinrib82}, and
the constant vorticity lines in turbulent vorticity fields~\cite{Bernard06}.
Empirical and numerical studies of isoheight lines show that they are
usually scale invariant~\cite{Mandelbrot83,Kondev95} and that their
fractal dimension $d_f$ depends on the long-range correlations, 
quantified by the Hurst exponent $H$~\cite{Schrenk13}. 

One solid framework to study fractal curves in two-dimensions is the
Schramm-Loewner Evolution ($SLE$) theory~\cite{Schramm00}. Accordingly,
an $SLE$ curve can be mapped onto a one-dimensional Brownian motion of a
single parameter in the continuum limit. Establishing that isoheight
lines are $SLE$ would allow us to generate an ensemble of such curves by
simply solving a stochastic differential equation, without generating
the entire landscape, and to extend established results from $SLE$ to
isoheight lines.

We analyze the zero isoheight lines of random surfaces of different $H$
generated with free boundary conditions on a triangular lattice of size
$L_x \times L_y$ with $L_x>L_y$. We consider \textit{chordal} curves
that are non-self-touching curves growing to infinity in the upper
half-plane $\mathbb{H}$, starting at the origin. From the Riemann
mapping theorem, there is a unique conformal map $g_t$ that iteratively
maps the complement of such a curve in the upper half-plane $\mathbb{H}$
back onto $\mathbb{H}$. This map satisfies the Loewner differential
equation,
\begin{equation}
\label{eq::Loewner_Eq}
\frac{\partial g_t(z)}{\partial t}=\frac{2}{g_t(z)-\xi_t},
\end{equation}
with $g_0(z)=z$ and $\xi_t$ a real continuous function, called the
driving function. If a curve is $SLE_\kappa$ then
$\xi_t=\sqrt{\kappa}B_t$, where $B_t$ is a standard one-dimensional
Brownian motion and $\kappa$ is the diffusion constant. We show that,
while for $H\leq0$ the statistics of $\xi_t$ is consistent with a
Brownian motion, for $H>0$, $\xi_t$ is not Markovian and therefore $SLE$
cannot be established.

We generate random landscapes on a triangular lattice, by assigning to
each lattice site $\mathbf{x}=(x,y)$ a random height $h(\mathbf{x})$,
imposing long-range correlations with the following spectrum,
\begin{equation}
\label{eq::autocorrelation_function}
S(\mathbf{q})\sim|\mathbf{q}|^{-\beta_c} \ \ ,
\end{equation}
where $\beta_c=2(H+1)$ and $H$ is the Hurst exponent. For that, we use
the Fourier Filtering Method~\cite{Prakash92,Makse96,Ahrens11} where,
\begin{equation} \label{eq::Fourier_Filtering_Method}
h\left(\mathbf{x}\right)=\tilde{\mathcal{F}}\left(\sqrt{S(\mathbf{q})}u(\mathbf{q})\right),
\end{equation}
where $\tilde{\mathcal{F}}$ denotes the inverse Fourier transform and
the $u(\mathbf{q})$ are independent complex Gaussian random variables of
mean zero and unitary variance satisfying
$u(\mathbf{-q})=\overline{u(\mathbf{q})}$. With this scheme, one
recovers uncorrelated landscapes for $H=-1$ and the discrete Gaussian
Free Field (GFF) for $H=0$ \cite{Lodhia14}. 

To obtain the zero isoheight lines, we extract the set of bonds on the
dual lattice separating the sites of negative height from the positive
ones.  We then focus on their accessible perimeter obtained by moving
along the isoheight line and shortcutting distances equal to the lattice
unit of the dual lattice, i.e., $\sqrt{3}/3$ lattice units of the
triangular lattice~\cite{Schrenk13}. Note that, using the formalism of
ranked surfaces, one can show that, for $H\leq0$, these lines correspond
to the accessible perimeter of a correlated percolation cluster at the
percolation threshold~\cite{Schrenk12,Schrenk13}. The accessible
perimeters of the zero isoheight lines on the triangular lattice in the
case of $H=-1$ and $H=0$ are analytically tractable and they have been
proven to be $SLE_\frac{8}{3}$ and $SLE_4$,
respectively~\cite{Schramm00,Smirnov01,Schramm09}. 

\textbf{SLE and fractal dimension}. As proven by
Beffara~\cite{Beffara08}, $SLE_\kappa$ curves are fractals of a fractal
dimension $d_f$ that is related to the diffusion coefficient $\kappa$
by,
\begin{equation} \label{eq::Fractal_Dim_kappa}
d_f=\min\left(2,1+\frac{\kappa}{8}\right).
\end{equation}
The fractal dimension $d_f$ of the accessible perimeter of the isoheight
lines for different values of $H$ was numerically estimated and even a
conjecture was proposed for its dependence on
$H$~\cite{Kondev95,Schrenk13}.  Using Eq.~(\ref{eq::Fractal_Dim_kappa}),
this gives a first estimate for the expected values of $\kappa$ if the
curves are $SLE$, see Table~\ref{tab::corr_perc_kappa}.  To verify if
$SLE$ can be established we will compare the $\kappa$ values calculated
from $d_f$ with estimates obtained with two indirect methods, the
winding angle and the left-passage probability, and with the one
obtained from the direct $SLE$ mapping. 

\begin{figure}[t]
\begin{center}
\includegraphics[width=\columnwidth]{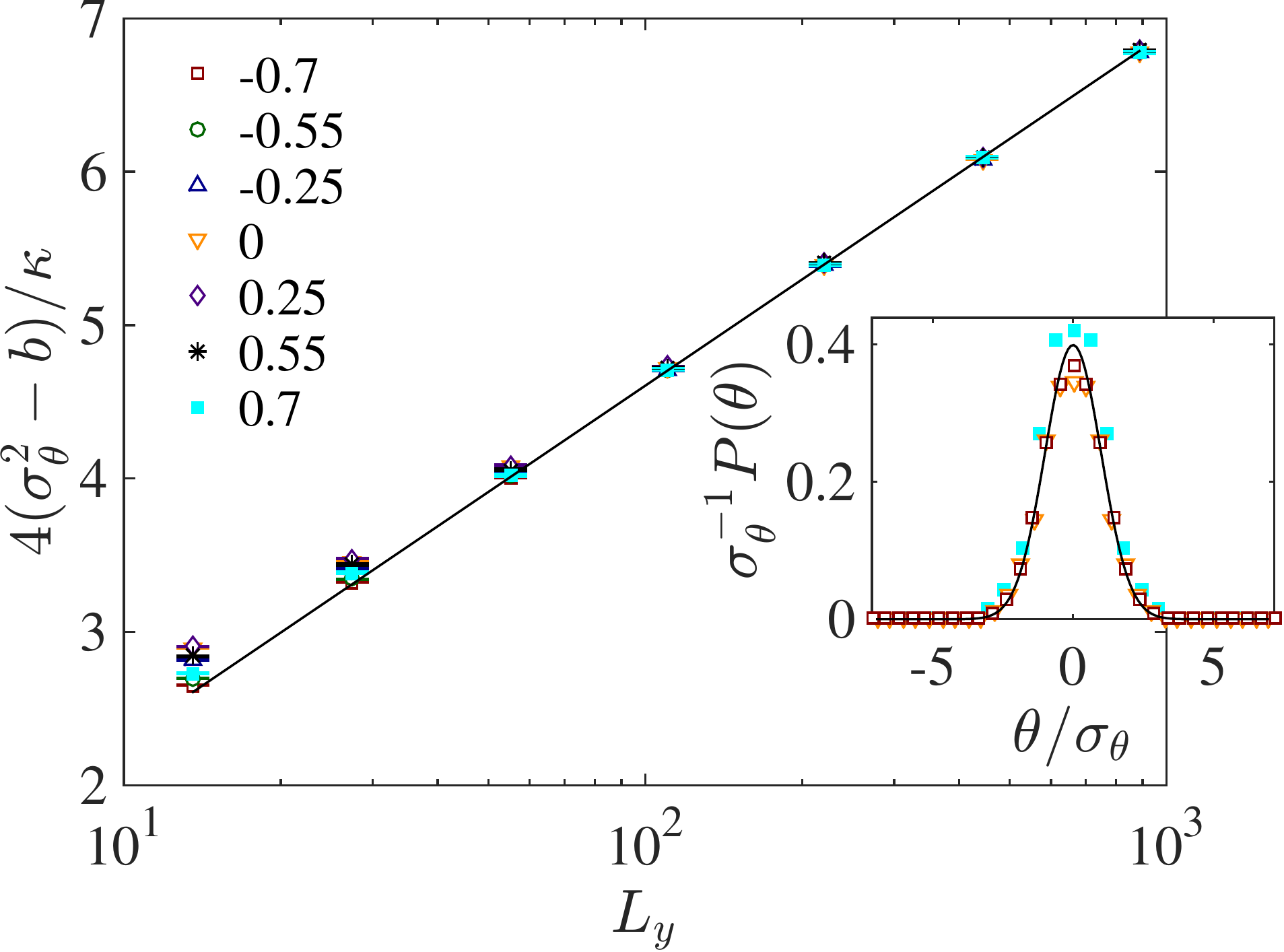}
\end{center}
\caption{\label{fig::corr_perc_windingAngle_all} (color online) Rescaled
variance of the winding angle for different Hurst exponents
$H=-0.7,-0.55,-0.25,0,0.25,0.55,0.7$. The black solid line represents
the relation $\frac{\sigma_{\theta}^2-b}{\kappa/4}=\ln L_y$. In the
inset, the rescaled probability distributions are plotted and compared
to a normal distribution for $H=-0.7,0,0.7$.}
\end{figure}

\textbf{Winding angle}. The winding angle of $SLE_\kappa$ curves follows
a Gaussian distribution and the variance scales with $\kappa$ as,
\begin{equation} \label{eq::winding_angle}
\langle \theta^2 \rangle -\langle \theta \rangle ^2= \sigma_{\theta}^2=b
+ \frac{\kappa}{4}\ln(L_y),
\end{equation}
where $b$ is a constant and $L_y$ is the vertical lattice size
\cite{Duplantier88, Schramm00, Wieland03}. To verify this relation for
each path, we consider the discrete set of points $z_i$ of the path. The
winding angle $\theta_i$ at each point $z_i$ can be computed iteratively
as $\theta_{i+1}=\theta_i+\alpha_i$, where $\alpha_i$ is the turning
angle between two consecutive points $z_i$ and $z_{i+1}$ on the
path. Duplantier and Saleur computed the probability distribution of the
winding angle for random curves using conformal invariance and Coulomb
gas techniques \cite{Duplantier88}. $\kappa/4$ corresponds to the slope
of $\sigma_{\theta}^2$ against $\ln(L_y)$, see
Fig.~\ref{fig::corr_perc_windingAngle_all}. The estimates of $\kappa$
are displayed in Table~\ref{tab::corr_perc_kappa}.

For values near $H=0$ and $H=1$, one has less precision on the results,
as the system is strongly influenced by finite-size effects, see
e.g. Ref.~\cite{Schrenk13}. The results we obtain from the winding angle
measurement are, within error bars, in agreement with the previous
estimates from the fractal dimension of the accessible perimeter of zero
isoheight lines. Indeed, Eq.~(\ref{eq::winding_angle}) gives insights
into the conformal invariance of the problem \cite{Boffeta08}. Our
results in Fig.~\ref{fig::corr_perc_windingAngle_all} give some
numerical indication that the accessible zero isoheight lines display
conformal invariance, which is a prerequisite for $SLE$.

\textbf{Left-Passage Probability}. As we simulate the curves in a
bounded rectangular domain, we map conformally the isoheight lines into
the upper half-plane, using an inverse Schwarz-Christoffel
transformation \cite{Pose14}, to obtain the \textit{chordal} curve,
which splits the system into two sides.  For chordal $SLE$ curves,
Schramm has computed the probability $P_{\kappa}(\phi)$ that a given
point $z=R e^{i\phi}$ in the upper half-plane $\mathbb{H}$ is on the
right-hand side of the curve \cite{Schramm01}.  This probability only
depends on $\phi$ and is given by Schramm's formula
\begin{equation} \label{eq::lpp_Schramm_formula}
P_{\kappa}(\phi)=\frac{1}{2}+\frac{\Gamma\left(4/\kappa\right)}{\sqrt{\pi}\Gamma\left(\frac{8-\kappa}{2\kappa}\right)}\cot(\phi)_2F_1\left(\frac{1}{2},\frac{4}{\kappa},
\frac{3}{2}, -\cot(\phi)^2\right),
\end{equation}
where $\Gamma$ is the Gamma function and $_2F_1$ is the Gauss
hypergeometric function \cite{Schrenk15}. This probability is known as
left-passage probability. 

We define a set of sample points $S$ in $\mathbb{H}$ for which we
measure the left-passage probability in order to compare it to the
values predicted by Schramm's formula (\ref{eq::lpp_Schramm_formula}).
To estimate $\kappa$, we minimize the mean square deviation $Q(\kappa)$
between the computed and predicted probabilities,
\begin{equation} \label{eq::lpp_quality_fit}
Q\left(\kappa\right)=\frac{1}{|S|}\sum_{z\in S}
\left[P(z)-P_{\kappa}(\phi)\right]^2, 
\end{equation}
where $\phi=\arg(z)$, $|S|$ is the cardinality of the set $S$, and
$P(z)$ the measured left-passage probability at $z$. The
estimated value of $\kappa$ corresponds to the point where the minimum
of $Q(\kappa)$ is observed.

\begin{figure}[t]
\begin{center}
\includegraphics[width=\columnwidth]{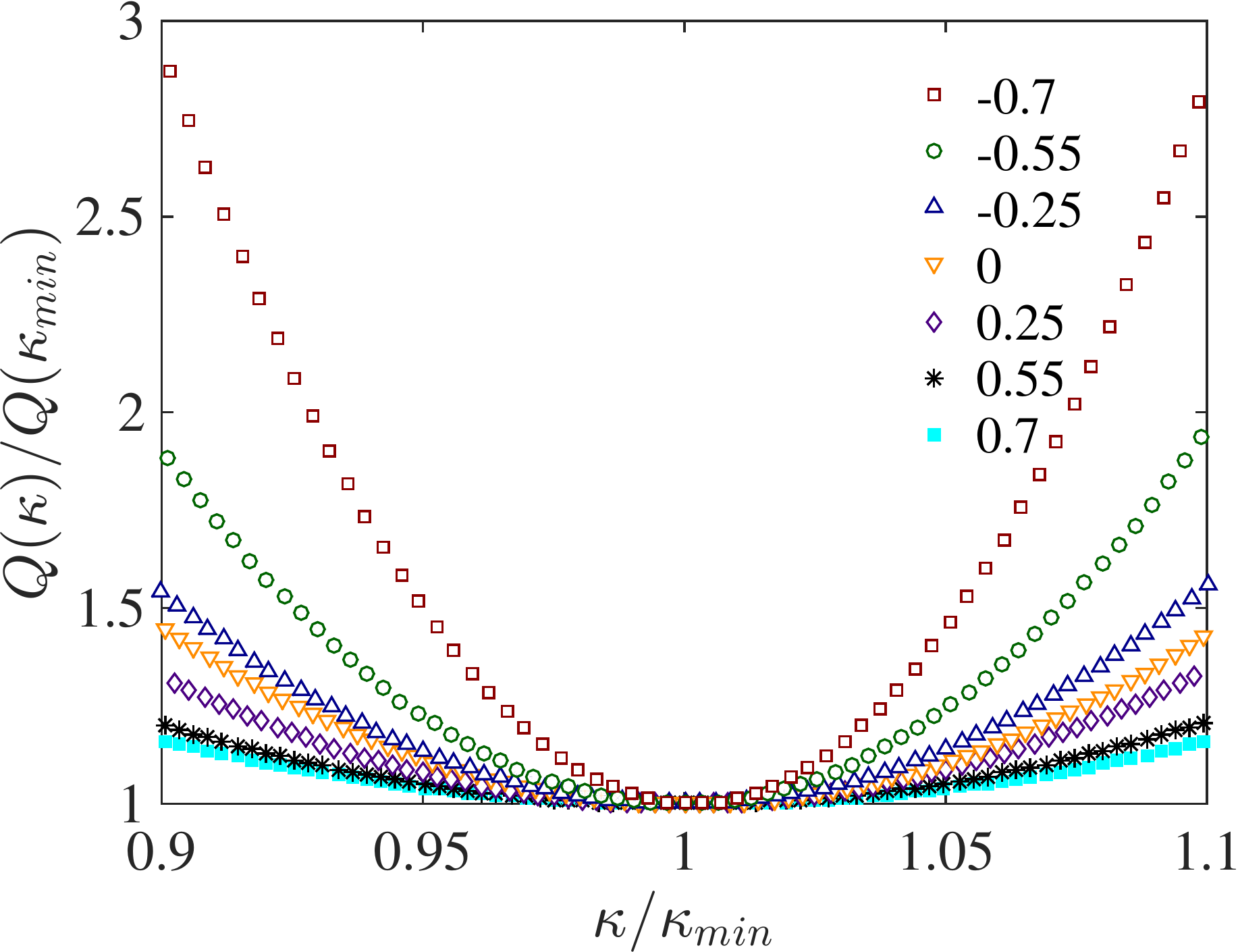}
\end{center}
\caption{\label{fig::corr_perc_lpp_all} (color online) Measured rescaled
mean square deviation $Q(\kappa)/Q(\kappa_{min})$ as a function of
$\kappa/\kappa_{min}$ with $\kappa_{min}$ the value of $\kappa$ where
the minimum of $Q(\kappa)$ is attained, for different Hurst exponents
$H=-0.7,-0.55,-0.25,0,0.25,0.55,0.7$.  We chose $50^2$ points in the
range $[-0.025L_x,0.025L_x]\times [0.15L_y,0.25L_y]$ with $L_y=1024$ and
$L_x=8L_y$, which are then mapped to the upper half-plane through an
inverse Schwarz-Christoffel transformation~\cite{Pose14}.}
\end{figure}

As shown in Fig.~\ref{fig::corr_perc_lpp_all}, for $H<0$, the minimum of
$Q$ is less pronounced for higher values of $H$, as it is expected for
functions of the form (\ref{eq::lpp_Schramm_formula}) with values of
$\kappa$ increasing towards four. As summarized in
Table~\ref{tab::corr_perc_kappa}, the estimated values of $\kappa$
obtained for negative $H$ are consistent with the ones predicted from
$d_f$ and confirmed by the winding angle analysis.  However, for $H>0$,
this is not the case. The obtained $\kappa$ values do not significantly
depend on $H$ and are consistently higher than the ones obtained from
$d_f$. This result suggest that for $H>0$ the curves are not $SLE$.

\textbf{Direct SLE}. To further test if the curves are $SLE$,
one has to check that the statistics of the driving function $\xi_t$ is
consistent with a one-dimensional Brownian motion with variance $\kappa
t$. This can be done by solving Eq.~(\ref{eq::Loewner_Eq}) numerically.
In order to do so, we use the so-called vertical slit map algorithm,
where one considers the driving function to be constant over small time
intervals $\delta t$. Making this approximation, one can solve
Eq.~(\ref{eq::Loewner_Eq}) to obtain the following slit map equation
\cite{Kennedy09,Cardy05},
\begin{equation} \label{eq::slit_map_Eq}
g_t(z)=\xi_t+\sqrt{\left(z-\xi_t\right)^2+4\delta t}.
\end{equation}
At $t=0$, one considers the initial curve consisting of the points
$\{z_{0}^{0}=0, \ldots , z_N^{0}=z_N\}$, and sets the driving function
to be $\xi_0=0$. Then at each iteration $i=1, \ldots,N$, we apply the
conformal map $g_{t_i}$ to the remaining points $z_j^{i-1}$ of the
curve, for $j=i, \ldots, N$ to obtain the new mapped curve. One gets a
new set of points $z_{j+1}^i=g_{t_i}(z_{j+1}^{i-1})$ for $j=i,\ldots,
N-1$ shorter by one point, by mapping $z_{i}^{i-1}$ to the real axis.
For that, in Eq.~(\ref{eq::slit_map_Eq}) we set $\xi_{t_i}=\mbox{Re} \{
z_i^{i-1} \}$ and $\delta t_i=t_i-t_{i-1}=\left(\mbox{Im} \{z_i^{i-1}\}
\right)^2/4$, where $\mbox{Re}\{\}$ and $\mbox{Im}\{\}$ are the real and
imaginary parts, respectively.

\begin{figure}[t]
\begin{center}
\includegraphics[width=\columnwidth]{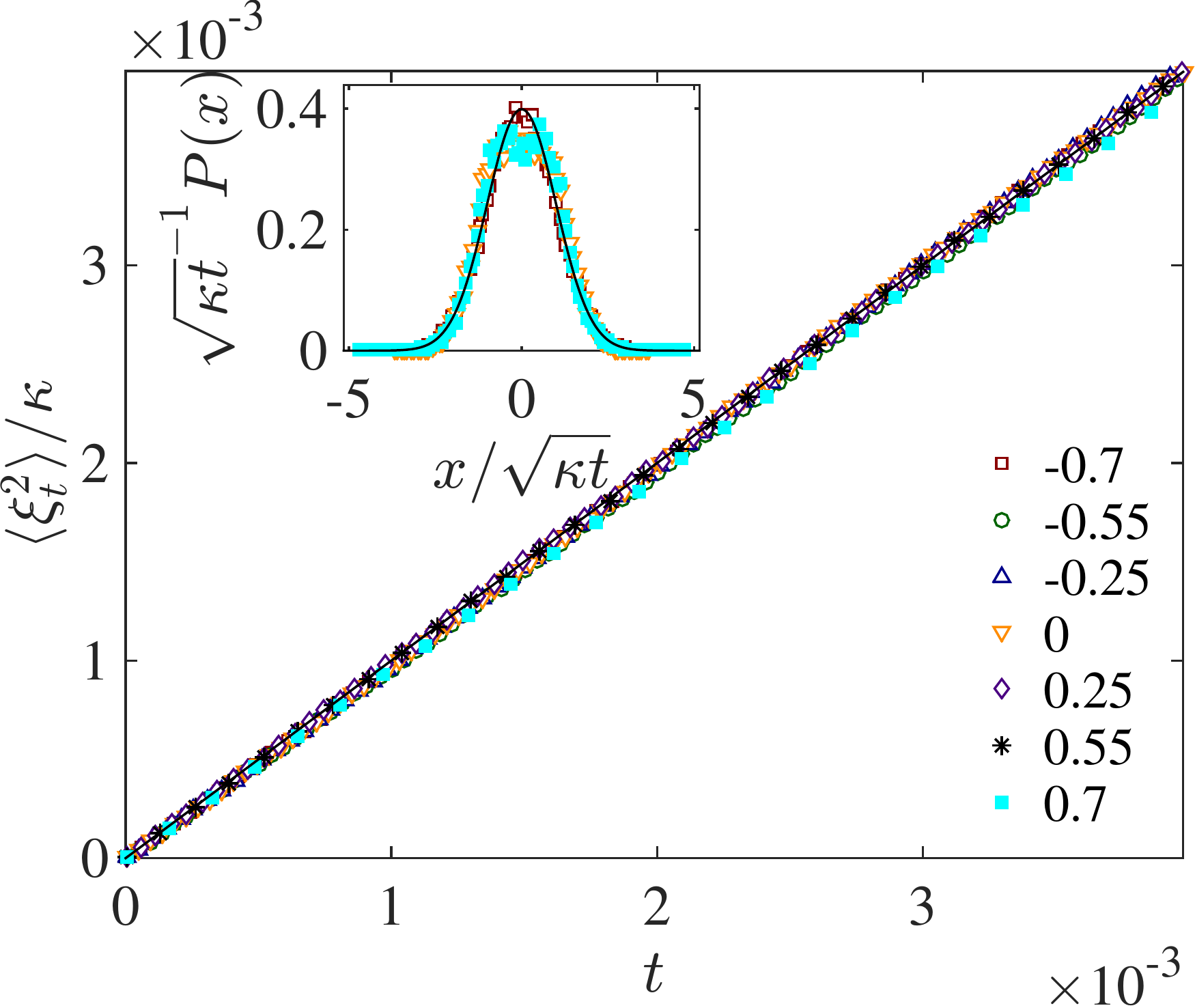}
\end{center}
\caption{\label{fig::corr_perc_dSLE_all} (color online) Rescaled
variance of the driving functions for different values of
$H=-0.7,-0.55,-0.25,0,0.25,0.55,0.7$. In the inset, we present the
rescaled probability distributions of the driving functions and compare
them to a Gaussian distribution for $H=-0.7,0,0.7$.  }
\end{figure}
We extracted the driving function of all paths and computed the
diffusion coefficient $\kappa$, see values in
Table~\ref{tab::corr_perc_kappa}, from the variance of the driving
function and tested its Gaussian distribution at a fixed Loewner time $t$,
see Fig.~\ref{fig::corr_perc_dSLE_all}. We find a linear scaling of the
variance of the driving function with the Loewner time. However, only
for $H\leq0$, the estimated values of $\kappa$ are consistent with the
ones predicted from $d_f$. For $H>0$, we obtain values of $\kappa$
significantly higher than the ones expected from $d_f$. In fact, the value of
$\kappa$ even increases with $H$, instead of decreasing as expected.

\begin{figure}
\includegraphics[width=\columnwidth]{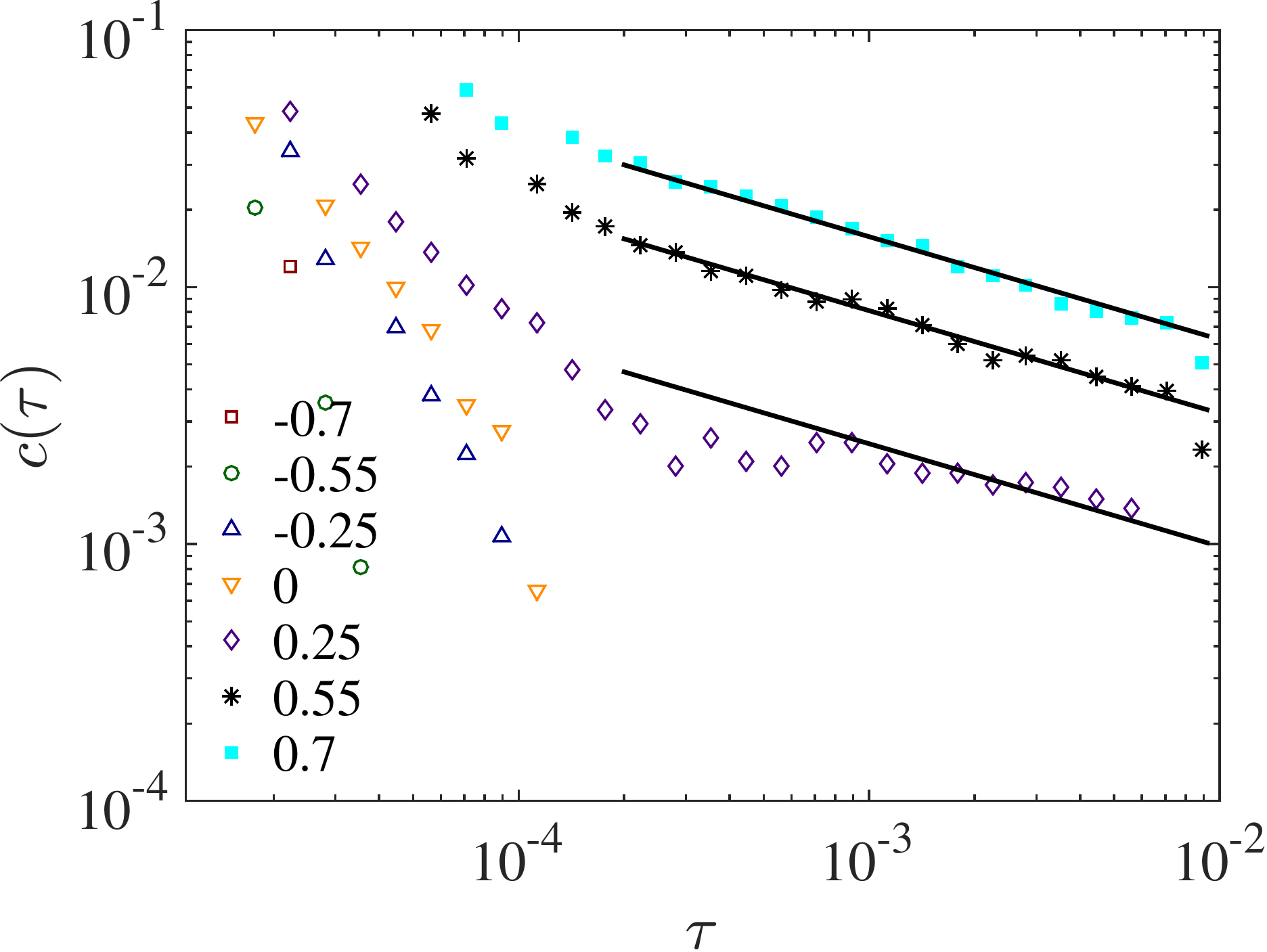}
\caption{\label{fig::markov_prop} Auto-correlation function $c(\tau)$ of
the increments for different values of
$H=-0.7,-0.55,-0.25,0,0.25,0.55,0.7$, averaged over $50$ time steps. The
solid lines are power laws of exponent $-0.4$.}
\end{figure}

We also test the Markovian property of the driving function by computing
the auto-correlation $c(\tau)=\langle c(t,\tau)\rangle_t$ of the
increments, with
\begin{equation}
c(t,\tau)=\frac{\langle \delta \xi_{t+\tau} \delta \xi_t\rangle-\langle
\delta \xi_{t+\tau}\rangle \langle \delta \xi_t
\rangle}{\sqrt{\left(\langle \delta \xi_{t+\tau}^2 \rangle - \langle
\delta \xi_{t+\tau}\rangle^2\right)\left( \langle \delta \xi_t^2 \rangle
- \langle \delta\xi_t \rangle^2\right)}},
\end{equation} as shown in Fig.~\ref{fig::markov_prop}, where
$\delta\xi_t$ is the change in the driving function at time $t$. For
$H\leq0$, the correlation function drops to zero after few time steps,
as expected for a Brownian motion. For $H>0$ we observe that the
correlation function decays as a power law and thus the driving function
is not Markovian, as it shows persistence in the increments. This
explains why, though the variance of the driving function is a linear
function of time, the statistics of the isoheight lines for $H>0$ are
not consistent with $SLE$.

\begin{figure}[t]
\begin{center}
\includegraphics[width=\columnwidth]{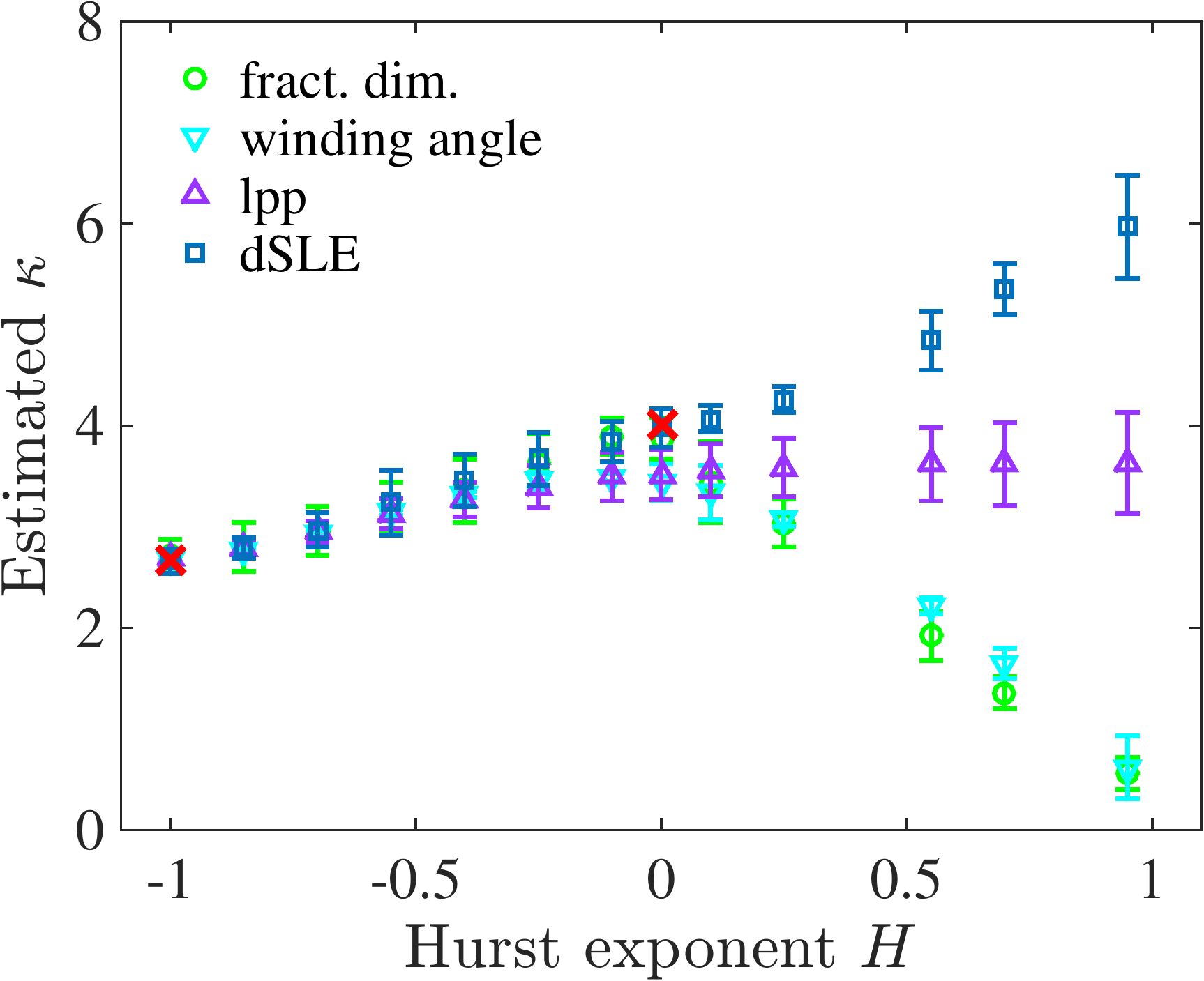}
\end{center}
\caption{\label{fig::corr_perc_final_plot_all} (color online) Estimated
diffusion coefficients $\kappa$ from the fractal dimension, the winding angle, the left-passage probability (lpp), and the
direct $SLE$ methods (dSLE). The red crosses correspond to the rigorous
results.}
\end{figure}

\textbf{Conclusion}.  We numerically showed that the statistics of the
accessible perimeter of the zero isoheight lines of long-range
correlated landscapes are consistent with $SLE$ only for $-1 \leq H \leq
0$. For this range, results for the fractal dimension, winding angle and
direct $SLE$ are in agreement within error bars, see
Fig.~\ref{fig::corr_perc_final_plot_all}.  This means that one can
describe these curves with a Brownian motion parameterized by a
diffusivity $\kappa$.  In the two analytic limits $H=-1$ and $H=0$, the
accessible perimeters are $SLE_{8/3}$ and $SLE_{4}$ for $H=-1$ and
$H=0$, respectively, results that we recover in our numerical analysis.
To our knowledge, this is the first time that for an entire range of
values of the Hurst exponent $H$, a family of curves coupled to random
landscapes is shown to be consistent with $SLE$. This gives new insight in the field of
fractional Gaussian Fields in two dimensions~\cite{Lodhia14}, what might
be helpful for understanding correlated landscapes from a theoretical
point of view. One might also wonder if these isoheight lines
can be related to some random walk process, as in the case of
uncorrelated random landscapes and the Gaussian Free Field. For example, the
isoheight lines for $H=-1$ and $H=0$ are two specific cases of the
overruled harmonic walker \cite{Celani09}. 

For positive $H$, we show that the driving function also scales linearly
in time but it is not Markovian, a necessary condition to establish
$SLE$. The numerical data suggest that the auto-correlation function
scales as a power law. Credidio~\textit{et al.} showed that if the
driving function is a stochastic process with anomalous diffusion, the
generated tracers are anisotropic~\cite{Credidio15}. In the same spirit,
it would be of interest to systematically study the statistics of random
curves generated from a driving function with persistence and compare
them to isoheight lines.

\begin{table}[t]
\begin{center}
\caption{\label{tab::corr_perc_kappa} Diffusion coefficient $\kappa$
computed from the fractal dimension $\kappa_{\mathrm{frac}}$ using data
from \cite{Schrenk13} for $H\leq0$ and obtained numerically for $H>0$
using the yardstick method, from the winding angle $\kappa_{\theta}$,
from the left-passage probability $\kappa_{\mathrm{LPP}}$ and the direct
$SLE$ method $\kappa_{\mathrm{dSLE}}$, for the different values of the
Hurst exponent $H$.}
\begin{ruledtabular}
\begin{tabular}{ccccc}
  $H$ & $\kappa_{\mathrm{frac}}$ & $\kappa_{\theta}$ &
$\kappa_{\mathrm{LPP}}$ & $\kappa_{\mathrm{dSLE}}$ \\
  \hline
  $-1$ & $2.76 \pm 0.16$ & $2.66 \pm 0.01$ & $2.69 \pm 0.08$ & $2.66 \pm
0.12$ \\
  $-0.85$ & $2.80 \pm 0.24$ & $2.76 \pm 0.02$ & $2.80 \pm 0.07$ & $2.79
\pm 0.10$ \\
  $-0.7$ & $2.96 \pm 0.24$ & $2.94 \pm 0.03$ & $2.95 \pm 0.11$ & $2.97
\pm 0.17$ \\
  $-0.55$ & $3.20 \pm 0.24$ & $3.14 \pm 0.03$ & $3.13 \pm 0.15$ & $3.29
\pm 0.32$ \\
  $-0.4$ & $3.35 \pm 0.31$ & $3.32 \pm 0.03$ & $3.27 \pm 0.17$ & $3.46
\pm 0.26$ \\
  $-0.25$ & $3.64 \pm 0.28$ & $3.45 \pm 0.09$ & $3.40 \pm 0.21$ & $3.67
\pm 0.26$ \\
  $-0.1$ & $3.90 \pm 0.18$ & $3.49 \pm 0.04$ & $3.50 \pm 0.24$ & $3.84
\pm 0.20$ \\
  $0$ & $3.88 \pm 0.20$ & $3.44 \pm 0.18$ & $3.52 \pm 0.25$ & $3.98 \pm
0.19$ \\
  $0.1$ & $3.44 \pm 0.40$ & $3.44 \pm 0.27$ & $3.56 \pm 0.26$ & $4.07
\pm 0.13$ \\
  $0.25$ & $3.04 \pm 0.24$ & $3.07 \pm 0.07$ & $3.59 \pm 0.29$ & $4.26
\pm 0.13$ \\
  $0.55$ & $1.92 \pm 0.24$ & $2.22 \pm 0.08$ & $3.62 \pm 0.36$ & $4.84
\pm 0.29$ \\
  $0.7$ & $1.36 \pm 0.16$ & $1.65 \pm 0.16$ & $3.62 \pm 0.41$ & $5.35
\pm 0.25$ \\
  $0.95$ & $0.56 \pm 0.16$ & $0.62 \pm 0.31$ & $3.63 \pm 0.50$ & $5.97
\pm 0.51$ \\
\end{tabular}
\end{ruledtabular}
\end{center}
\end{table}

This Letter also opens the possibility of applying $SLE$ to the study of
landscapes with negative Hurst exponent $H$. As we have shown, many
systems can be considered from the point of view of a landscape, where
isoheights are of relevance. Especially, it has been shown that
zero vorticity isolines in two-dimensional turbulence are $SLE$
\cite{Bernard06}. There have been also attempts to extend this result to
isolines in a generalized Navier-Stokes equation \cite{Falkovich10} to
study the conformal invariance of a larger class of turbulence problems.
It would be interesting to see if a relation between this problem and
our results can be drawn for the accessible perimeters of these contour
lines. 

\begin{acknowledgments}
The authors would like to thank W. Werner for helpful discussions. We
acknowledge financial support from the European Research Council (ERC)
Advanced Grant 319968-FlowCCS, support by the Swiss National Science
Foundation under Grant No. P2EZP2-152188, and the Portuguese Foundation
for Science and Technology (FCT) under contracts no. IF/00255/2013,
UID/FIS/00618/2013, and EXCL/FIS-NAN/0083/2012.
\end{acknowledgments}

\bibliography{bibliography_SLE}

\begin{thebibliography}{26}%
\makeatletter
\providecommand \@ifxundefined [1]{%
 \@ifx{#1\undefined}
}%
\providecommand \@ifnum [1]{%
 \ifnum #1\expandafter \@firstoftwo
 \else \expandafter \@secondoftwo
 \fi
}%
\providecommand \@ifx [1]{%
 \ifx #1\expandafter \@firstoftwo
 \else \expandafter \@secondoftwo
 \fi
}%
\providecommand \natexlab [1]{#1}%
\providecommand \enquote  [1]{``#1''}%
\providecommand \bibnamefont  [1]{#1}%
\providecommand \bibfnamefont [1]{#1}%
\providecommand \citenamefont [1]{#1}%
\providecommand \href@noop [0]{\@secondoftwo}%
\providecommand \href [0]{\begingroup \@sanitize@url \@href}%
\providecommand \@href[1]{\@@startlink{#1}\@@href}%
\providecommand \@@href[1]{\endgroup#1\@@endlink}%
\providecommand \@sanitize@url [0]{\catcode `\\12\catcode `\$12\catcode
  `\&12\catcode `\#12\catcode `\^12\catcode `\_12\catcode `\%12\relax}%
\providecommand \@@startlink[1]{}%
\providecommand \@@endlink[0]{}%
\providecommand \url  [0]{\begingroup\@sanitize@url \@url }%
\providecommand \@url [1]{\endgroup\@href {#1}{\urlprefix }}%
\providecommand \urlprefix  [0]{URL }%
\providecommand \Eprint [0]{\href }%
\providecommand \doibase [0]{http://dx.doi.org/}%
\providecommand \selectlanguage [0]{\@gobble}%
\providecommand \bibinfo  [0]{\@secondoftwo}%
\providecommand \bibfield  [0]{\@secondoftwo}%
\providecommand \translation [1]{[#1]}%
\providecommand \BibitemOpen [0]{}%
\providecommand \bibitemStop [0]{}%
\providecommand \bibitemNoStop [0]{.\EOS\space}%
\providecommand \EOS [0]{\spacefactor3000\relax}%
\providecommand \BibitemShut  [1]{\csname bibitem#1\endcsname}%
\let\auto@bib@innerbib\@empty
\bibitem [{\citenamefont {Boffetta}\ \emph {et~al.}(2008)\citenamefont
  {Boffetta}, \citenamefont {Celani}, \citenamefont {Dezzani},\ and\
  \citenamefont {Seminara}}]{Boffeta08}%
  \BibitemOpen
  \bibfield  {author} {\bibinfo {author} {\bibfnamefont {G.}~\bibnamefont
  {Boffetta}}, \bibinfo {author} {\bibfnamefont {A.}~\bibnamefont {Celani}},
  \bibinfo {author} {\bibfnamefont {D.}~\bibnamefont {Dezzani}}, \ and\
  \bibinfo {author} {\bibfnamefont {A.}~\bibnamefont {Seminara}},\ }\href@noop
  {} {\bibfield  {journal} {\bibinfo  {journal} {Geophys. Res. Lett.}\ }\textbf
  {\bibinfo {volume} {35}},\ \bibinfo {pages} {L03615} (\bibinfo {year}
  {2008})}\BibitemShut {NoStop}%
\bibitem [{\citenamefont {Ziman}(1968)}]{Ziman68}%
  \BibitemOpen
  \bibfield  {author} {\bibinfo {author} {\bibfnamefont {J.~M.}\ \bibnamefont
  {Ziman}},\ }\href@noop {} {\bibfield  {journal} {\bibinfo  {journal} {J.
  Phys. C}\ }\textbf {\bibinfo {volume} {1}},\ \bibinfo {pages} {1532}
  (\bibinfo {year} {1968})}\BibitemShut {NoStop}%
\bibitem [{\citenamefont {Weinrib}(1982)}]{Weinrib82}%
  \BibitemOpen
  \bibfield  {author} {\bibinfo {author} {\bibfnamefont {A.}~\bibnamefont
  {Weinrib}},\ }\href@noop {} {\bibfield  {journal} {\bibinfo  {journal} {Phys.
  Rev. B}\ }\textbf {\bibinfo {volume} {26}},\ \bibinfo {pages} {1352}
  (\bibinfo {year} {1982})}\BibitemShut {NoStop}%
\bibitem [{\citenamefont {Bernard}\ \emph {et~al.}(2006)\citenamefont
  {Bernard}, \citenamefont {Boffetta}, \citenamefont {Celani},\ and\
  \citenamefont {Falkovich}}]{Bernard06}%
  \BibitemOpen
  \bibfield  {author} {\bibinfo {author} {\bibfnamefont {D.}~\bibnamefont
  {Bernard}}, \bibinfo {author} {\bibfnamefont {G.}~\bibnamefont {Boffetta}},
  \bibinfo {author} {\bibfnamefont {A.}~\bibnamefont {Celani}}, \ and\ \bibinfo
  {author} {\bibfnamefont {G.}~\bibnamefont {Falkovich}},\ }\href@noop {}
  {\bibfield  {journal} {\bibinfo  {journal} {Nat. Phys.}\ }\textbf {\bibinfo
  {volume} {2}},\ \bibinfo {pages} {124} (\bibinfo {year} {2006})}\BibitemShut
  {NoStop}%
\bibitem [{\citenamefont {Mandelbrot}(1983)}]{Mandelbrot83}%
  \BibitemOpen
  \bibfield  {author} {\bibinfo {author} {\bibfnamefont {B.~B.}\ \bibnamefont
  {Mandelbrot}},\ }\href@noop {} {\emph {\bibinfo {title} {The Fractal Geometry
  of Nature}}}\ (\bibinfo  {publisher} {Freeman},\ \bibinfo {address} {New
  York},\ \bibinfo {year} {1983})\BibitemShut {NoStop}%
\bibitem [{\citenamefont {Kondev}\ and\ \citenamefont
  {Henley}(1995)}]{Kondev95}%
  \BibitemOpen
  \bibfield  {author} {\bibinfo {author} {\bibfnamefont {J.}~\bibnamefont
  {Kondev}}\ and\ \bibinfo {author} {\bibfnamefont {C.~L.}\ \bibnamefont
  {Henley}},\ }\href@noop {} {\bibfield  {journal} {\bibinfo  {journal} {Phys.
  Rev. Lett.}\ }\textbf {\bibinfo {volume} {74}},\ \bibinfo {pages} {4580}
  (\bibinfo {year} {1995})}\BibitemShut {NoStop}%
\bibitem [{\citenamefont {Schrenk}\ \emph {et~al.}(2013)\citenamefont
  {Schrenk}, \citenamefont {Pos\'e}, \citenamefont {Kranz}, \citenamefont
  {\mbox{van Kessenich}}, \citenamefont {Ara\'ujo},\ and\ \citenamefont
  {Herrmann}}]{Schrenk13}%
  \BibitemOpen
  \bibfield  {author} {\bibinfo {author} {\bibfnamefont {K.~J.}\ \bibnamefont
  {Schrenk}}, \bibinfo {author} {\bibfnamefont {N.}~\bibnamefont {Pos\'e}},
  \bibinfo {author} {\bibfnamefont {J.~J.}\ \bibnamefont {Kranz}}, \bibinfo
  {author} {\bibfnamefont {L.~V.~M.}\ \bibnamefont {\mbox{van Kessenich}}},
  \bibinfo {author} {\bibfnamefont {N.~A.~M.}\ \bibnamefont {Ara\'ujo}}, \ and\
  \bibinfo {author} {\bibfnamefont {H.~J.}\ \bibnamefont {Herrmann}},\
  }\href@noop {} {\bibfield  {journal} {\bibinfo  {journal} {Phys. Rev. E}\
  }\textbf {\bibinfo {volume} {88}},\ \bibinfo {pages} {052102} (\bibinfo
  {year} {2013})}\BibitemShut {NoStop}%
\bibitem [{\citenamefont {Schramm}(2000)}]{Schramm00}%
  \BibitemOpen
  \bibfield  {author} {\bibinfo {author} {\bibfnamefont {O.}~\bibnamefont
  {Schramm}},\ }\href@noop {} {\bibfield  {journal} {\bibinfo  {journal} {Isr.
  J. Math.}\ }\textbf {\bibinfo {volume} {118}},\ \bibinfo {pages} {221}
  (\bibinfo {year} {2000})}\BibitemShut {NoStop}%
\bibitem [{\citenamefont {Prakash}\ \emph {et~al.}(1992)\citenamefont
  {Prakash}, \citenamefont {Havlin}, \citenamefont {Schwartz},\ and\
  \citenamefont {Stanley}}]{Prakash92}%
  \BibitemOpen
  \bibfield  {author} {\bibinfo {author} {\bibfnamefont {S.}~\bibnamefont
  {Prakash}}, \bibinfo {author} {\bibfnamefont {S.}~\bibnamefont {Havlin}},
  \bibinfo {author} {\bibfnamefont {M.}~\bibnamefont {Schwartz}}, \ and\
  \bibinfo {author} {\bibfnamefont {H.~E.}\ \bibnamefont {Stanley}},\
  }\href@noop {} {\bibfield  {journal} {\bibinfo  {journal} {Phys. Rev. A}\
  }\textbf {\bibinfo {volume} {46}},\ \bibinfo {pages} {R1724} (\bibinfo {year}
  {1992})}\BibitemShut {NoStop}%
\bibitem [{\citenamefont {Makse}\ \emph {et~al.}(1996)\citenamefont {Makse},
  \citenamefont {Havlin}, \citenamefont {Schwartz},\ and\ \citenamefont
  {Stanley}}]{Makse96}%
  \BibitemOpen
  \bibfield  {author} {\bibinfo {author} {\bibfnamefont {H.~A.}\ \bibnamefont
  {Makse}}, \bibinfo {author} {\bibfnamefont {S.}~\bibnamefont {Havlin}},
  \bibinfo {author} {\bibfnamefont {M.}~\bibnamefont {Schwartz}}, \ and\
  \bibinfo {author} {\bibfnamefont {H.~E.}\ \bibnamefont {Stanley}},\
  }\href@noop {} {\bibfield  {journal} {\bibinfo  {journal} {Phys. Rev. E}\
  }\textbf {\bibinfo {volume} {53}},\ \bibinfo {pages} {5445} (\bibinfo {year}
  {1996})}\BibitemShut {NoStop}%
\bibitem [{\citenamefont {Ahrens}\ and\ \citenamefont
  {Hartmann}(2011)}]{Ahrens11}%
  \BibitemOpen
  \bibfield  {author} {\bibinfo {author} {\bibfnamefont {B.}~\bibnamefont
  {Ahrens}}\ and\ \bibinfo {author} {\bibfnamefont {A.~K.}\ \bibnamefont
  {Hartmann}},\ }\href@noop {} {\bibfield  {journal} {\bibinfo  {journal}
  {Phys. Rev. B}\ }\textbf {\bibinfo {volume} {84}},\ \bibinfo {pages} {144202}
  (\bibinfo {year} {2011})}\BibitemShut {NoStop}%
\bibitem [{\citenamefont {Lodhia}\ \emph {et~al.}()\citenamefont {Lodhia},
  \citenamefont {Sheffield}, \citenamefont {Sun},\ and\ \citenamefont
  {Watson}}]{Lodhia14}%
  \BibitemOpen
  \bibfield  {author} {\bibinfo {author} {\bibfnamefont {A.}~\bibnamefont
  {Lodhia}}, \bibinfo {author} {\bibfnamefont {S.}~\bibnamefont {Sheffield}},
  \bibinfo {author} {\bibfnamefont {X.}~\bibnamefont {Sun}}, \ and\ \bibinfo
  {author} {\bibfnamefont {S.~S.}\ \bibnamefont {Watson}},\ }\href@noop {} {\
  }\Eprint {http://arxiv.org/abs/arXiv:1407.5598v1} {arXiv:1407.5598v1}
  \BibitemShut {NoStop}%
\bibitem [{\citenamefont {Schrenk}\ \emph {et~al.}(2012)\citenamefont
  {Schrenk}, \citenamefont {Ara\'ujo}, \citenamefont {{Andrade Jr.}},\ and\
  \citenamefont {Herrmann}}]{Schrenk12}%
  \BibitemOpen
  \bibfield  {author} {\bibinfo {author} {\bibfnamefont {K.~J.}\ \bibnamefont
  {Schrenk}}, \bibinfo {author} {\bibfnamefont {N.~A.~M.}\ \bibnamefont
  {Ara\'ujo}}, \bibinfo {author} {\bibfnamefont {J.~S.}\ \bibnamefont {{Andrade
  Jr.}}}, \ and\ \bibinfo {author} {\bibfnamefont {H.~J.}\ \bibnamefont
  {Herrmann}},\ }\href {\doibase 10.1038/srep00348} {\bibfield  {journal}
  {\bibinfo  {journal} {Sci. Rep.}\ }\textbf {\bibinfo {volume} {2}},\ \bibinfo
  {pages} {348} (\bibinfo {year} {2012})}\BibitemShut {NoStop}%
\bibitem [{\citenamefont {Smirnov}(2001)}]{Smirnov01}%
  \BibitemOpen
  \bibfield  {author} {\bibinfo {author} {\bibfnamefont {S.}~\bibnamefont
  {Smirnov}},\ }\href@noop {} {\bibfield  {journal} {\bibinfo  {journal} {C. R.
  Acad. Sci. Paris I}\ }\textbf {\bibinfo {volume} {333}},\ \bibinfo {pages}
  {239} (\bibinfo {year} {2001})}\BibitemShut {NoStop}%
\bibitem [{\citenamefont {Schramm}\ and\ \citenamefont
  {Sheffield}(2009)}]{Schramm09}%
  \BibitemOpen
  \bibfield  {author} {\bibinfo {author} {\bibfnamefont {O.}~\bibnamefont
  {Schramm}}\ and\ \bibinfo {author} {\bibfnamefont {S.}~\bibnamefont
  {Sheffield}},\ }\href@noop {} {\bibfield  {journal} {\bibinfo  {journal}
  {Acta. Math.}\ }\textbf {\bibinfo {volume} {202}},\ \bibinfo {pages} {21}
  (\bibinfo {year} {2009})}\BibitemShut {NoStop}%
\bibitem [{\citenamefont {Beffara}(2008)}]{Beffara08}%
  \BibitemOpen
  \bibfield  {author} {\bibinfo {author} {\bibfnamefont {V.}~\bibnamefont
  {Beffara}},\ }\href@noop {} {\bibfield  {journal} {\bibinfo  {journal} {Ann.
  Probab.}\ }\textbf {\bibinfo {volume} {36}},\ \bibinfo {pages} {1421}
  (\bibinfo {year} {2008})}\BibitemShut {NoStop}%
\bibitem [{\citenamefont {Duplantier}\ and\ \citenamefont
  {Saleur}(1988)}]{Duplantier88}%
  \BibitemOpen
  \bibfield  {author} {\bibinfo {author} {\bibfnamefont {B.}~\bibnamefont
  {Duplantier}}\ and\ \bibinfo {author} {\bibfnamefont {H.}~\bibnamefont
  {Saleur}},\ }\href@noop {} {\bibfield  {journal} {\bibinfo  {journal} {Phys.
  Rev. Lett.}\ }\textbf {\bibinfo {volume} {60}},\ \bibinfo {pages} {2343}
  (\bibinfo {year} {1988})}\BibitemShut {NoStop}%
\bibitem [{\citenamefont {Wieland}\ and\ \citenamefont
  {Wilson}(2003)}]{Wieland03}%
  \BibitemOpen
  \bibfield  {author} {\bibinfo {author} {\bibfnamefont {B.}~\bibnamefont
  {Wieland}}\ and\ \bibinfo {author} {\bibfnamefont {D.~B.}\ \bibnamefont
  {Wilson}},\ }\href@noop {} {\bibfield  {journal} {\bibinfo  {journal} {Phys.
  Rev. E}\ }\textbf {\bibinfo {volume} {68}},\ \bibinfo {pages} {056101}
  (\bibinfo {year} {2003})}\BibitemShut {NoStop}%
\bibitem [{\citenamefont {Pos\'e}\ \emph {et~al.}(2014)\citenamefont {Pos\'e},
  \citenamefont {Schrenk}, \citenamefont {Ara\'ujo},\ and\ \citenamefont
  {Herrmann}}]{Pose14}%
  \BibitemOpen
  \bibfield  {author} {\bibinfo {author} {\bibfnamefont {N.}~\bibnamefont
  {Pos\'e}}, \bibinfo {author} {\bibfnamefont {K.~J.}\ \bibnamefont {Schrenk}},
  \bibinfo {author} {\bibfnamefont {N.~A.~M.}\ \bibnamefont {Ara\'ujo}}, \ and\
  \bibinfo {author} {\bibfnamefont {H.~J.}\ \bibnamefont {Herrmann}},\
  }\href@noop {} {\bibfield  {journal} {\bibinfo  {journal} {Sci. Rep.}\
  }\textbf {\bibinfo {volume} {4}},\ \bibinfo {pages} {5495} (\bibinfo {year}
  {2014})}\BibitemShut {NoStop}%
\bibitem [{\citenamefont {Schramm}(2001)}]{Schramm01}%
  \BibitemOpen
  \bibfield  {author} {\bibinfo {author} {\bibfnamefont {O.}~\bibnamefont
  {Schramm}},\ }\href {\doibase 10.1214/ECP.v6-1041} {\bibfield  {journal}
  {\bibinfo  {journal} {Electron. Commun. Prob.}\ }\textbf {\bibinfo {volume}
  {6}},\ \bibinfo {pages} {115} (\bibinfo {year} {2001})}\BibitemShut {NoStop}%
\bibitem [{\citenamefont {Schrenk}\ and\ \citenamefont
  {Stevenson}()}]{Schrenk15}%
  \BibitemOpen
  \bibfield  {author} {\bibinfo {author} {\bibfnamefont {K.~J.}\ \bibnamefont
  {Schrenk}}\ and\ \bibinfo {author} {\bibfnamefont {J.~D.}\ \bibnamefont
  {Stevenson}},\ }\href@noop {} {}\Eprint
  {http://arxiv.org/abs/arXiv:1502.05624} {arXiv:1502.05624} \BibitemShut
  {NoStop}%
\bibitem [{\citenamefont {Kennedy}(2009)}]{Kennedy09}%
  \BibitemOpen
  \bibfield  {author} {\bibinfo {author} {\bibfnamefont {T.}~\bibnamefont
  {Kennedy}},\ }\href@noop {} {\bibfield  {journal} {\bibinfo  {journal} {J.
  Stat. Phys.}\ }\textbf {\bibinfo {volume} {137}},\ \bibinfo {pages} {839}
  (\bibinfo {year} {2009})}\BibitemShut {NoStop}%
\bibitem [{\citenamefont {Cardy}(2005)}]{Cardy05}%
  \BibitemOpen
  \bibfield  {author} {\bibinfo {author} {\bibfnamefont {J.}~\bibnamefont
  {Cardy}},\ }\href@noop {} {\bibfield  {journal} {\bibinfo  {journal} {Ann.
  Phys. (N.Y.)}\ }\textbf {\bibinfo {volume} {318}},\ \bibinfo {pages} {81}
  (\bibinfo {year} {2005})}\BibitemShut {NoStop}%
\bibitem [{\citenamefont {Celani}\ \emph {et~al.}(2009)\citenamefont {Celani},
  \citenamefont {Mazzino},\ and\ \citenamefont {Tizzi}}]{Celani09}%
  \BibitemOpen
  \bibfield  {author} {\bibinfo {author} {\bibfnamefont {A.}~\bibnamefont
  {Celani}}, \bibinfo {author} {\bibfnamefont {A.}~\bibnamefont {Mazzino}}, \
  and\ \bibinfo {author} {\bibfnamefont {M.}~\bibnamefont {Tizzi}},\
  }\href@noop {} {\bibfield  {journal} {\bibinfo  {journal} {J. Stat. Mech.}\
  ,\ \bibinfo {pages} {P12011}} (\bibinfo {year} {2009})}\BibitemShut {NoStop}%
\bibitem [{\citenamefont {Credidio}\ \emph {et~al.}()\citenamefont {Credidio},
  \citenamefont {Moreira}, \citenamefont {Herrmann},\ and\ \citenamefont
  {\mbox{Andrade Jr.}}}]{Credidio15}%
  \BibitemOpen
  \bibfield  {author} {\bibinfo {author} {\bibfnamefont {H.~F.}\ \bibnamefont
  {Credidio}}, \bibinfo {author} {\bibfnamefont {A.~A.}\ \bibnamefont
  {Moreira}}, \bibinfo {author} {\bibfnamefont {H.~J.}\ \bibnamefont
  {Herrmann}}, \ and\ \bibinfo {author} {\bibfnamefont {J.~S.}\ \bibnamefont
  {\mbox{Andrade Jr.}}},\ }\href@noop {} {}\Eprint
  {http://arxiv.org/abs/arXiv:1308.5692} {arXiv:1308.5692} \BibitemShut
  {NoStop}%
\bibitem [{\citenamefont {Falkovich}\ and\ \citenamefont
  {Musacchio}()}]{Falkovich10}%
  \BibitemOpen
  \bibfield  {author} {\bibinfo {author} {\bibfnamefont {G.}~\bibnamefont
  {Falkovich}}\ and\ \bibinfo {author} {\bibfnamefont {S.}~\bibnamefont
  {Musacchio}},\ }\href@noop {} {}\Eprint
  {http://arxiv.org/abs/arXiv:1012.3868v1} {arXiv:1012.3868v1} \BibitemShut
  {NoStop}%
\end{thebibliography}%

\end{document}